\documentclass{pasj00}

\draft

\begin{document}

\SetRunningHead{Y. Takeda et al.}{S and Zn abundances of red giants}
\Received{2016/06/17}
\Accepted{2016/07/07}

\title{Sulfur and zinc abundances of red giant stars\thanks{
Based on data collected at Okayama Astrophysical Observatory
of NAOJ (Okayama, Japan).}
}

%

\author{
Yoichi \textsc{Takeda,}\altaffilmark{1,2}
Masashi \textsc{Omiya,}\altaffilmark{1}
Hiroki \textsc{Harakawa,}\altaffilmark{1}
 and
Bun'ei \textsc{Sato}\altaffilmark{3}
}

\altaffiltext{1}{National Astronomical Observatory, 2-21-1 Osawa, 
Mitaka, Tokyo 181-8588}
\email{takeda.yoichi@nao.ac.jp}
\altaffiltext{2}{SOKENDAI, The Graduate University for Advanced Studies, 
2-21-1 Osawa, Mitaka, Tokyo 181-8588}
\altaffiltext{3}{Tokyo Institute of Technology, 2-12-1 Ookayama, 
Meguro-ku, Tokyo 152-8550}

\KeyWords{stars: abundances --- stars: atmospheres ---  
stars: evolution --- stars: late-type 
}

\maketitle

\begin{abstract}
Sulfur and zinc are chemically volatile elements, which play significant 
roles as depletion-free tracers in studying galactic chemical evolution.
However, regarding red giants having evolved off the main sequence, reliable 
abundance determinations of S and Zn seem to be difficult despite that a few 
studies have been reported so far. 
Given this situation, we tried to establish the abundances of these elements for 
an extensive sample of 239 field  GK giants ($-0.8 \ltsim$~[Fe/H]~$\ltsim +0.2$), 
by applying the spectrum-fitting technique to S~{\sc i} 8694--5, S~{\sc i} 6757, 
and Zn~{\sc i} 6362 lines and by taking into account the non-LTE effect.
Besides, similar abundance analysis was done for 160 FGK dwarfs to be used for comparison.
The non-LTE corrections for the S and Zn abundances derived from these lines turned out
$\ltsim$~0.1(--0.2)~dex for most cases and not very significant. It revealed that the S~{\sc i} 6757 
feature is more reliable as an abundance indicator than S~{\sc i} 8694--5 for the 
case of red giants, because the latter suffers blending of unidentified lines. 
The finally resulting [S/Fe]--[Fe/H] and [Zn/Fe]--[Fe/H] relations for GK giants 
were confirmed to be in good agreement with those for FGK dwarfs, indicating 
that S and Zn abundances of red giants are reliably determinable from the 
S~{\sc i} 6757 and Zn~{\sc 6362} lines. Accordingly, not only main-sequence stars
but also evolved red giant stars are usable for tracing the chemical evolution history of 
S and Zn in the regime of disk metallicity by using these lines.  
\end{abstract}

%


\section{Introduction}

Sulfur and zinc are counted as important elements in stellar spectroscopy. 
Owing to their ``volatile'' characteristics, they are hardly affected 
by condensation process (unlike refractory elements such as Fe, 
which may suffer gas--dust depletion). For this reason, S ($\alpha$ group) 
and Zn (Fe-group element) are regarded to play significant roles in galactic
chemical evolution, since they may be used as depletion-independent tracers
of nucleosynthesis history. 
Accordingly, not a few spectroscopic investigations on the behavior of 
[S/Fe] and [Zn/Fe] in stars of various metallicities have been done toward 
clarifying the chemical enrichment history of these elements in the Galaxy 
(see, e.g., Takada-Hidai et al. 2005 for S, Caffau et al. 2005 for S, Saito et al. 
2009 for Zn, Takeda et al. 2005c for S and Zn, and the references therein). 

However, regarding stars of Galactic disk population, unevolved dwarfs of 
F--G type have been primarily used for such studies of S and Zn abundances, 
while few reliable abundance results of these elements are available for 
red giants (evolved low--to-intermediate-mass stars), despite that
several extensive abundance studies on a large sample of field giants 
have been reported in the past decade:\\
--- S and Zn were out of the target elements in Takeda, Sato, and Murata's 
(2008; hereinafter referred to as Paper~I) analysis for 322 G--K giants 
due to the limited wavelength range of the spectra.\\
--- In Luck and Heiter's (2007) spectroscopic study of 298 field giants,
S and Zn abundances were determined only for 10 and 6 stars, respectively,
which were not sufficient for establishing the trends of [S/Fe] or [Zn/Fe].\\
--- Luck (2015) recently reported the abundance results for a very extensive 
sample of 1133 field G--K giants. While S and Zn abundances were determined 
for most stars, the dispersions of [Element/Fe] ratios are unusually large
($\sim \pm 0.4$~dex) only for these two elements (cf. Figure~5 therein), 
which means that they are unreliable. Actually, as noted by Luck (2015) 
himself, [S/Fe] systematically grows from $\sim 0$ 
($T_{\rm eff} \sim 5500$~K) to $\sim 2$ ($T_{\rm eff} \sim 4000$~K), 
manifestly indicating an involvement of appreciable blending (with unknown 
lines) becoming progressively important toward 
later spectral types (see Figure~6 therein). 
Regarding Zn, Luck (2015) similarly remarked that ``The zinc lines are 
badly blended, especially at temperatures below 4800K, resulting in 
abundances of poor quality.'' \\
--- Very recently, Maldonaldo and Villaver (2016) published the abundances
of 16 elements (including S and Zn) for a large sample of evolved stars
and compared those of unevolved main-sequence stars. As seen from their
Fig. A.1, [S/Fe] results for evolved stars are systematically higher than 
those for main-sequence stars, and the dispersion of [Zn/Fe] for giants 
is appreciably larger than that for dwarfs, which implies that larger errors
may be involved in their abundances of these elements for red giants.\\
--- In Jofr\'{e} et al.'s (2015) comparison of their abundance results
for giant stars with literature studies (cf. Fig.~7 therein), Zn abundances 
were compared with those of Maldonaldo, Villaver, and Eiroa (2013); but 
the consistency was not necessarily satisfactory.

It is thus evident that abundance studies of these two elements for evolved 
red giant stars are still at quite unsatisfactory level, which are yet 
to be conducted by a careful analysis of appropriate spectral lines.
Takeda et al. (2015; hereinafter referred to as Paper~II) investigated
the oxygen abundances (along with those of C and Na) for 239 late G
and early K giants by using the $R$- and $I$-band spectra (up to 
$\sim 8800$~$\rm\AA$) newly obtained in the 2012--2013 period.
This was originally intended to supplement/revise the previous work 
of Paper~I, which was based on old spectra covering only 
the yellow--orange region ($\sim$~5000--6200~$\rm\AA$).
Since these new longer-wavelength region spectra comprises important spectral
lines usable for determinations of S and Zn abundances (S~{\sc i} 8693--5,
S~{\sc i} 6757, and Zn~{\sc i} 6362), we decided to make use these data
and determine the S and Zn abundances of 239 red giants by using 
the spectrum-fitting method while properly including the non-LTE effect.

Accordingly, we would like to examine in this investigation whether abundances 
of S and Zn can be reliably determined for low-gravity giants of G--K type. 
Toward this purpose, the S and Zn abundances of unevolved 160 dwarfs (including 
subgiants) of late-F through early-K type were also derived in essentially 
the same manner, which we intend to use as reference samples for comparing
the abundance trends. That is, given that both S and Zn should not suffer any 
surface abundance change (unlike the abundances of some light elements such 
as Li or C, which are affected by evolution-induced mixing of nuclear-processed
materials), we may expect that essentially the same [S/Fe]--[Fe/H] or 
[Zn/Fe]--[Fe/H] relation would be obtained for both GK giants and FGK dwarfs. 
If such a consistency could be confirmed, we may state that trustworthy 
abundances of these elements are established also for giants. 

Since abundance studies of two samples (GK giants and FGK dwarfs) are involved 
in this investigation based on different data sets by making use of our previous 
results, we firstly outline the adopted observational data, the atmospheric 
parameters, and the method of line-stregth measurement (to be described 
in section~2 and section~3) in table~1 for the reader's convenience.

The remainder of this article is organized as follows: The abundance 
determination procedures (spectrum-fitting analysis, equivalent-width 
derivation, and derivation of non-LTE abundances) for our 239 program stars 
of GK giants are described in section~2. Section~3 is devoted to a brief 
description of S and Zn abundances for the reference sample of 160 FGK dwarfs.
The resulting abundances along with the related quantities are discussed 
in section~4, where the [Fe/H]-dependent trends for giants and dwarfs are
compared with each other. The conclusions are summarized in section~5.   

\section{S and Zn abundances of GK giants}

The high-dispersion spectra of 239 G--K giants used for this study are the same 
as adopted in Paper~II (see section~2 therein), which were obtained in 2012--2013 
by using the HIDES at the coud\'{e} focus of the 188 cm reflector at Okayama 
Astrophysical Observatory. Regarding the atmospheric parameters [the effective 
temperature ($T_{\rm eff}$), surface gravity ($\log g$), Fe abundance relative 
to the Sun ([Fe/H]), and microturbulence ($v_{\rm t}$)], we exclusively 
adopted those spectroscopically determined in Paper~I unchanged (cf. subsection~3.1
therein). The atmospheric model of each star corresponding to these parameters 
was generated by interpolating Kurucz's (1993) ATLAS9 model grid as in Paper~I.

Our abundance determination closely follows the procedure adopted in Paper~II 
(see section~3 therein for more details), which consists of three consecutive steps: 
(i) synthetic spectrum fitting to derive provisional abundances for the important 
elements in the relevant spectrum range, (ii) inverse calculation of the equivalent 
width for the line (or the blended feature) of S or Zn in question by using the 
fitting-based abundance solutions, and (iii) analysis of such established equivalent 
widths to derive the final abundances or abundance errors due to uncertainties 
in atmospheric parameters.
 
Given that new $R$- and $I$-band spectra (up to $\sim 8800$~$\rm\AA$) are available
for our targets, we selected three wavelength regions for abundance determinations
of S and Zn: (a) 8691--8696~$\rm\AA$ region comprising the S~{\sc i} 8693--5 triplet 
lines (along with Ti~{\sc i}~8692.33 and Nd~{\sc ii}~8685.17), 
(b)  6756--6758.1~$\rm\AA$ region comprising the blended feature of S~{\sc i} 6757 
lines (along with Fe~{\sc i}~6756.56 and Co~{\sc i}~6757.71, and 
(c) 6362--6365~$\rm\AA$ region comprising the Zn~{\sc i} 6362.34 line 
(along with [O~{\sc i}] + CN 6363 feature) which we already analyzed in Paper~II for
oxygen abundance determination.
The adopted atomic data of the relevant S and Zn lines are summarized in table~2.\footnote{
Regarding the atomic data of other spectral lines included in these spectral ranges, 
which are needed for spectrum synthesis calculations, we basically adopted Kurucz and 
Bell's (1995) compilation. However, we applied rough corrections to the original $\log gf$ 
values given therein for the following lines, in order to accomplish a better fit:
Nd~{\sc ii} 8695.170 (by +1.0~dex), 
Ni~{\sc i} 6757.714 (by +1.0~dex),
Si~{\sc i} 6758.005 (by $-1.0$~dex; i.e., reduced to a negligible level), and
Ni~{\sc i} 6758.061 (by $-1.0$~dex; i.e., reduced to a negligible level).}

The convergence of the fitting solutions turned out fairly successful  
for most of the cases. How the theoretical spectrum for the converged 
solutions fits well with the observed spectrum for each of the 239 GK giants 
is displayed in figure 1 (8691--8696~$\rm\AA$ fitting, where S, Ti, and Nd abundances
were varied) and figure 2 (6756--6758.1~$\rm\AA$ fitting, where S, Fe, and Co 
abundances were varied). See Fig.~4 in Paper~II for the case of 6362-6365~$\rm\AA$ 
fitting where O, Fe, Zn, and CN abundances were varied.

Then, based on these fitting-based abundances, we computed the equivalent widths for 
the S~{\sc i} 8694.63 line ($W_{8695}$; the strongest component of the triplet), 
the S~{\sc i} 6757 feature ($W_{6757}$; comprising three components), and the 
Zn~{\sc i} 6362.34 line ($W_{6362}$), from which the non-LTE abundances 
($A$)\footnote{We use $A$ to denote the logarithmic elemental
abundance (often expressed as $\log\epsilon$) relative to hydrogen.
That is, if number density in the atmosphere for element X and H is 
$N$(X) and $N$(H), respectively, $A$(X) is defined as 
$A$(X) $\equiv$ $\log N$(X) $-$ $\log N$(H) + 12.}  
and the relevant non-LTE corrections ($\Delta$) were derived,
where the non-LTE effect for S ad Zn lines was taken into account based on
the grid of non-LTE departure coefficients calculated by Takeda et al. (2005c)
with the standard treatment of H~{\sc i} collision cross section (i.e., $h =0$).
The resulting equivalent widths and  non-LTE abundances as well as corrections
for the 239 giants are given in tableE1.txt (on-line material). 

In order to estimate abundance errors due to ambiguities in atmospheric parameters,
we computed the changes in the S and Zn abundances while perturbing the standard 
atmospheric parameters interchangeably by $\pm 100$~K in $T_{\rm eff}$, $\pm 0.2$~dex 
in $\log g$, and $\pm 0.2$~km~s$^{-1}$ in $\xi$, as done in Paper~II. The mean 
values of the resulting abundance changes averaged over 239 stars are summarized 
in table~3, We can see from this table that the typical extents of such uncertainties 
are within $\ltsim 0.1$~dex and thus not so significant. 

\section{Comparison sample of FGK dwarfs}

Regarding the S and Zn abundances of 160 FGK dwarfs (including some subgiants), with 
which the abundance results of 239 giants are to be compared, essentially the same 
determination procedures as described in section~2 (i.e., non-LTE analysis of 
$W_{8695}$, $W_{6757}$, and $W_{6362}$) were adopted. Our analysis closely 
follows Takeda's (2007) study, which is based on the spectra published by 
Takeda et al. (2005a) along with the atmospheric parameters determined by 
Takeda et al. (2005b).

The $W_{8695}$ data (for S~{\sc i} 8694.63) were inversely derived by using the 
solutions of 8692.5--8695.5~$\rm\AA$ fitting done by Takeda (2007; cf. Fig.~4 therein),
while Takeda's (2007) directly measured $W_{6362}$ data were used for Zn~{\sc i} 6362.34.  
As to the S~{\sc i} 6757 line feature which was not analyzed by Takeda (2007), we newly 
carried out a spectrum-fitting analysis in the 6756--6758.1~$\rm\AA$ region and 
computed $W_{6757}$ for each star (as done for giants).
Figure~3 shows how the fitting between the observed and theoretical spectra 
has been accomplished for each of the 160 targets.

The equivalent widths and the resulting non-LTE abundances as well as corrections
for these 160 dwarfs(+subgiants) are presented in tableE2.txt (on-line material). 

\section{Results and discussion}

\subsection{Behaviors of abundances and related quantities}

We now examine the abundances as well as the related quantities resulting
from our analysis described in section~2 (giants) and section~3 (dwarfs).  

The strengths of S~{\sc i}~8695, S~{\sc i}~6757, and Zn~{\sc i}~6362 lines 
($W_{8695}$, $W_{6757}$, and $W_{6362}$) are plotted against $T_{\rm eff}$ and 
$\log g$ in figure~4, where how the non-LTE corrections ($\Delta$) depend 
upon $W$ is also shown.  We can recognize a tendency that S~{\sc i} lines
tend to be strengthened with an increase in $T_{\rm eff}$ or with a decrease 
in $\log g$, while such a trend is not clear for the case of Zn~{\sc i} 6362
line in dwarfs. 
The non-LTE corrections are negative ($\Delta < 0$) for most cases
(except for the Zn~{\sc i} 6362 line for dwarfs, where $\Delta$ tends to become
positive at the weak-line limit; cf. figure~4i)
and their extents ($|\Delta|$) show a tight correlation with $W$ 
(increasing with an increase in $W$).  Besides, since the non-LTE effect is
enhanced in the condition of lower gravity (lower density), the $|\Delta|$
values for giants are larger than those for dwarfs (when compared 
at the same $W$). Generally speaking, the non-LTE corrections are $\ltsim 0.1$~dex
for most cases (for all cases of S~{\sc i} 6757 and Zn~{\sc i} 6362) 
and not very significant, except for the case of especially 
stronger S~{\sc i} 8695 line corresponding to higher-luminosity giants of 
$\log g \ltsim 2$ where $|\Delta|$ becomes as large as $\sim$~0.2--0.3~dex
(cf. figure~4g).

Each of the non-LTE abundances [$A$(S)$_{8695}$, $A$(S)$_{6757}$, and 
$A$(Zn)$_{6362}$] are plotted against $T_{\rm eff}$ and $\log g$ in
figure~5, where $A$(S)$_{8695}$ vs. $A$(S)$_{6757}$ correlations
for giants and dwarfs are also shown.
We can see that the S abundances derived from S~{\sc i}~8695 and S~{\sc i}~6757 
lines are fairly good agreement for dwarfs (figure~5h) but not for giants
(figure~5g); i.e., the scatter is rather large and $A$(S)$_{8695}$ tends to be 
systematically larger than $A$(S)$_{6757}$ in the regime of high $A$(S). 
Regarding giants, we can recognize that the fit for S~{\sc i}~8693--5 lines 
(figure~2) is apparently not so good as for S~{\sc i}~6757 feature (figure~3).
This suggests that blending effect due to other lines (not included in
our analysis; e.g., molecular lines) may possibly take effect for the 
former S~{\sc i}~8693--5 case especially at lower $T_{\rm eff}$ and 
higher [Fe/H], making $A$(S)$_{8695}$ more or less overestimated. 
Actually, a signature of overestimation for $A$(S)$_{8695}$
is also observed in the maximum abundance ($A^{\rm max}$) among
the sample. Since [Fe/H]$^{\rm max}$ is $\sim$~+0.4--0.5 for dwarfs
(Takeda 2007) and $\sim +0.2$ for giants (Paper~I), similar
difference should be observed also for $A^{\rm max}$ between dwarfs and 
giants. While such a discordance amounting 0.2--0.3~dex is surely recognized 
in $A$(S)$_{6757}$ (figure~5e) and $A$(Zn)$_{6362}$ (figure~5f),
we can not detect any difference in the maximum values of $A$(S)$_{8695}$ 
between dwarfs and giants in figure~5d, which implies that $A$(S)$_{8695}$
values of giants are likely to be overestimated at the high-metallicity regime 
(presumably due to appreciable blending effect). Accordingly, 
we may state that S abundances derived from S~{\sc i}~6757 are more 
reliable than those from S~{\sc i}~8695 as far as red giants are concerned,
because the latter is more vulnerable to blending effects resulting in 
appreciable overestimation. 

This consequence is further substantiated by figure~6, where the resulting values 
of $W$, $\Delta$, and [X/Fe]\footnote{As usual, [X/Fe] is defined as
[X/Fe] $\equiv$ [X/H] $-$ [Fe/H] = 
[$A_{*}$(X) $-$ $A_{\odot}$(X)] $-$ [$A_{*}$(Fe) $-$ $A_{\odot}$(Fe)],
where $A$(X) is the non-LTE abundance derived in this study and [Fe/H]
is the Fe abundance relative to the Sun determined in Paper~I (giants) and
Takeda et al. (2005b) (dwarfs). Here, the subscripts $*$ and $\odot$ denote 
``star'' and ``Sun'', respectively.} (X is S or Zn) for each line are plotted 
against [Fe/H] (metallicity).
We can see that [S/Fe] vs. [Fe/H] relation for 239 giants constructed 
from $A$(S)$_{8695}$ (filled symbols in figure~6g) is appreciably different from
that constructed based on $A$(S)$_{6757}$ (filled symbols in figure~6h) in the sense that 
the former shows a significantly larger scatter than the latter. This manifestly
indicates that $A$(S)$_{8695}$ is less reliable than $A$(S)$_{6757}$ and thus
should not be used for the case of red giants.\footnote{
Actually, this argument applies not only to giants but also to dwarfs,
since the dispersion of [S/Fe]$_{8695}^{\rm dwarfs}$ (figure~6g)
appears to be larger than that for [S/Fe]$_{6757}^{\rm dwarfs}$ 
(figure~6h), which suggests that the S~{\sc i} 6757 feature
is more reliable as S abundance indicator also for FGK dwarfs.
It was once argued by Takeda (2007) based on the analysis of S~{\sc i} 
8693--5 lines that the dispersion of [S/Fe] for 160 dwarfs is appreciably
larger than that of [Si/Fe] (cf. subsection~4.2 therein), in contrast 
to the result of Chen et al. (2002) who derived rather tight 
distributions for both [S/Fe] and [Si/Fe].
However, given the present result, Takeda's (2007) conclusion may 
be worth reconsideration, which may bave been attributed to
larger uncertainties in the S abundances derived therein.
}
Hereinafter we will exclusively adopt 
$A$(S)$_{6757}$ as the final sulfur abundances for both giants and dwarfs.
According to figure~6h and figure~6i, both giants (filled symbols) and dwarfs (open symbols)
appear to follow almost the same  (and reasonably tight) [S/Fe] vs. [Fe/H] (from 
S~{\sc i}~6757) and [Zn/Fe] vs. [Fe/H] (from Zn~{\sc i}~6362) relations. 
We will actually confirm this fact in subsection~4.3.

\subsection{Comparison of S and Zn abundances for giants with published results}

The [S/H] and [Zn/H] values (differential S or Zn abundances relative
to the Sun) for giants derived in this study are compared with the literature 
results in figure~7. 

Regarding sulfur, all the [S/H] values published by Luck (2015), Luck \& Wepfer (1995),
and Maldonado \& Villaver (2016) tend to be systematically larger (typically by several 
tenths dex) than our [S/H] derived from S~{\sc i}~6757, as shown in figure~7a, figure~7c,
and figure~7e, respectively.
Although  Luck (2015) as well as Luck and Wepfer (1995) did not explicitly describe 
the lines employed for their S abundance determination, we presume that they used 
any of the S~{\sc i} 8670.200, 8670.627, 8671.308, 8678.950, 8680.445, 8693.150, 
8693.958, and 8694.641 lines, as inferred from the line list presented in 
Luck and Lambert (1985; cf. Table~1 therein). It is thus likely that they derived 
overestimated S abundances, because near-infrared ($I$-band) lines tend to suffer 
blending effect as we confirmed for S~{\sc i}~8693--5 lines.
Meanwhile, Maldonado and Villaver (2016) used S~{\sc i} 6046.00, 6052.66, and 6757.17
lines as included in Table~3 of Maldonado et al. (2015). While S~{\sc i} 6757 line
among these is the same as we adopted, the other two S~{\sc i} 6046/6052 lines may have 
suffered some appreciable blending with unknown lines. In any case, since 
equivalent-width data of the adopted lines are not published in these three literature 
studies, we can not trace down the true reason for this discrepancy. 

As to zinc, we compared our [Z/H] results with those of five previous studies in figure~7.
It can be seen from figure~7b, figure~7d, and figure~7g that Luck (2015), 
Luck \& Wepfer (1995), and Jofr\'{e} et al. (2015) yielded systematically larger 
[Zn/H] values (by several tenths dex) than our results. The adopted Zn line 
is again not explicitly stated in the former two papers (though we presume they 
employed Zn~{\sc i} 6362 line), while Jofr\'{e} et al. (2015) used Zn~{\sc i} 4722.16 
and 4810.53 lines. Although the reason for this discrepancy is not clear,
their results may have been influenced by blending effect and overestimated.
On the other hand, the [Zn/H] values derived by Maldonado \& Villaver (2016) 
as well as Maldonado, Villaver, \& Eiroa (2013) are in agreement with
our results, as illustrated in figure~7f and figure~7h. Maldonado et al. (2013) 
used Zn~{\sc i} 4810 and 6362 lines (as explicitly stated in Sect. 2.5 therein).
Meanwhile, Maldonado and Villaver (2016) seem to have used only Zn~{\sc i} 6362 line,
since they say that their line list was taken from Maldonado et al. (2015).\footnote{
Maldonado et al. (2015) state in their Sect. 2.6 that the Zn line was 
taken from Table~1 of Takeda and Honda (2005). But this must be incorrect, 
since any Zn line is not included therein. Instead, we consider that they 
adopted the Zn~{\sc i} 6362 line from Table~4 of Ram\'{\i}rez, Mel\'{e}ndez, 
and Asplund (2014) which was also referenced by them.} The small changes between 
the results of their two papers may be mainly due to difference in the adopted Zn lines. 

\subsection{Trends of [S/Fe] and [Zn/Fe] for giants and dwarfs}

Finally, we compare the metallicity-dependent trend of [S/Fe] (based on
S~{\sc i} 6757 line) as well as [Zn/Fe] (based on Zn~{\sc i} 6362 line)
derived for giants with that of dwarfs, and examine whether they satisfactorily
match each other. This can be a valuable check for the validity of
our S and Zn abundances of red giants as described in section~1. 
  
The [X/Fe] vs. [Fe/H] diagrams (X is S or Zn) plotted for giants and dwarfs 
are shown in the left-side panels (a, b) of figure~8, where the mean 
$\langle$[X/Fe]$\rangle$ at each metallicity group (0.1~dex bin within 
$-0.4 \le$~[Fe/H]~$\le +0.2$) along with the distribution of
$\langle$[X/Fe]$\rangle_{\rm giants} - \langle$[X/Fe]$\rangle_{\rm dwarfs}$ 
are also presented in the corresponding right-side panels (c, d).
We can confirm from this figure that [S/Fe]--[Fe/H] as well as [Zn/Fe]--[Fe/H]
relations are essentially the same for both giants and dwarfs 
(the difference is only $\ltsim 0.05$~dex), which substantiates the reliability
of our S and Zn abundances derived for giants as well as dwarfs.

From the viewpoint of galactic chemical evolution, these similar runs of [S/Fe]
and [Zn/Fe] with [Fe/H] (i.e., mild enhancement from $\sim 0.0$ to $\sim +0.3$
as [Fe/H] is decreased from $\sim 0$ down to $\sim -1$) are almost consistent 
with previous observational studies
(e.g., Takada-Hidai et al. 2005 or Caffau et al. 2005 for S; 
Saito et al. 2009 for Zn) as well as theoretical predictions (e.g., 
Kobayashi et al. 2006; see also Kobayashi \& Nomoto 2009 for Zn).

To conclude, the consequence of this study indicates that S and Zn abundances of
evolved red giants (for which few credible results have been reported so far)
can be reliably determined by carefully analyzing the S~{\sc i} 6757 and Zn~{\sc i} 6362
lines. Accordingly, we can employ (not only dwarfs but also) red giants to investigate 
the chemical evolution of S and Zn in the Galaxy by using these lines,
though they are suitable only at the metallicity range of disk population 
(i.e. $-1 \ltsim$~[Fe/H]).  

\section{Summary and conclusion}

Since sulfur and zinc belong to chemically volatile species, they play 
important roles in stellar spectroscopy because of being usable as 
depletion-free tracers for studying the galactic chemical evolution.

However, regarding stars of disk population, spectroscopic studies 
of S and Zn abundances have been mostly directed to unevolved dwarfs.
As a matter of fact, despite that several extensive abundance studies of 
evolved red giants have been published in the past decade, reliable
abundance determinations for these two elements seem to be rare.

Given this situation, we decided to establish the abundances of S and Zn 
for 239 apparently bright GK giants (in the metallicity range of 
$-0.8 \ltsim$~[Fe/H]~$\ltsim +0.2$) by applying the spectrum-fitting 
technique to S~{\sc i} 8694--5, S~{\sc i} 6757, and Zn~{\sc 6362} lines,
where the non-LTE effect was also taken into account.
Besides, S and Zn abundances for 160 FGK dwarfs were also determined
in the same manner, which are to used for comparison.

The non-LTE corrections for the S and Zn abundances derived from these 
lines turned out mostly $\ltsim$~0.1~dex (though amounting up to 
$\sim$~0.2--0.3~dex in some exceptional cases of S~{\sc i} 8695) and not very significant. 
The S~{\sc i} 6757 feature was found to be more reliable as an abundance 
indicator than S~{\sc i} 8694--5 for the case of red giants, because 
the latter appears to suffers appreciable blending with unidentified lines. 

The finally obtained [S/Fe]--[Fe/H] and [Zn/Fe]--[Fe/H] relations 
for GK giants were confirmed to be in good agreement with those for 
FGK dwarfs, indicating that S and Zn abundances of red giants are 
reliably determinable from the S~{\sc i} 6757 and Zn~{\sc 6362} lines. 

Consequently, not only unevolved main-sequence stars but also evolved 
red giant stars can be exploited as probe to study the chemical evolution 
history of S and Zn in the regime of disk metallicity.  

\bigskip

Data reduction was in part carried out by using the common-use data analysis 
computer system at the Astronomy Data Center (ADC) of the National Astronomical 
Observatory of Japan.

\clearpage

\setcounter{table}{0}
\begin{table}[h]
\scriptsize
\caption{Brief outline of observational data, atmospheric parameters, 
and line-strength measurements.}
\begin{center}
\begin{tabular}
{ccc}\hline \hline
             &   239 GK giants      &     160 FGK dwarfs  \\
\hline
         \multicolumn{3}{c}{[Basic observational data]}\\
Observation years  &  2012--2013    &       2000--2003    \\
Wavelength range & 5100-8800~$\rm\AA$ & 5000--8800~$\rm\AA$ \\
            & (cf. Paper II) &    (cf. Takeda et al. 2005a)\\
Spectrum resolution & $\sim 67000$    &   $\sim 67000$  \\
\hline
         \multicolumn{3}{c}{[Atmospheric parameters]}\\
Method  &   Using FeI/FeII lines  &  Using FeI/FeII lines \\
        &   based on 2000--2005 data & based on the same 2000--2003 data \\
        &       (cf. Paper I)   &  (cf. Takeda et al. 2005b)\\
$T_{\rm eff}$ (K) &  $\sim$~4500--5600 &   $\sim$~5000--7000 \\
log g (cm~s$^{-2}$/dex)  &  $\sim$~1.4--3.5 &  $\sim$~3.1--4.9 \\
$v_{\rm t}$ (km~s$^{-1}$) & $\sim$~1.0--3.2 &  $\sim$~0.5--3.0 \\
$[$Fe/H$]$ (dex) & $\sim -0.8$ to $\sim +0.2$ & $\sim -1.3$ to $\sim +0.5$\\
\hline
         \multicolumn{3}{c}{[Line-strength evaluation]}\\
S~{\sc i} 8695 &  fitting (figure~1 of this paper) & fitting (Fig.~4 of Takeda 2007) \\
S~{\sc i} 6757 &  fitting (figure~2 of this paper) & fitting (figure~3 of this paper) \\
Zn~{\sc i} 6362 & fitting (Fig.~4 of Paper~II)  &  direct $W$ measurement (cf. Takeda 2007) \\
\hline
\end{tabular}
\end{center}
\end{table}

\setcounter{table}{1}
\begin{table}[h]
\scriptsize
\caption{Atomic data of the relevant S~{\sc i} and Zn~{\sc i} lines.}
\begin{center}
\begin{tabular}
{crccrccccc}\hline \hline
Species & Line & RMT & Multiplet & $\lambda$ ($\rm\AA)$ & $\chi_{\rm low}$ (eV) & $\log gf$ & Gammar & Gammas & Gammaw \\ \hline
S~{\sc i} &  8693 & 6 &   4p $^{5}{\rm P}_{3}$ -- 4d $^{5}{\rm D}_{2}^{\rm o}$ & 8693.137 & 7.870 & $-1.370$ & 7.62 & $-4.41$ & ($-7.30$) \\
S~{\sc i} &  8694 & 6 &   4p $^{5}{\rm P}_{3}$ -- 4d $^{5}{\rm D}_{3}^{\rm o}$ & 8693.931 & 7.870 & $-0.510$ & 7.62 & $-4.41$ & ($-7.30$) \\
S~{\sc i} &  8695 & 6 &   4p $^{5}{\rm P}_{3}$ -- 4d $^{5}{\rm D}_{4}^{\rm o}$ & 8694.626 & 7.870 & +0.080 & 7.62 & $-4.41$ & ($-7.30$) \\
S~{\sc i} &  6757 & 8 &  4p $^{5}{\rm P}_{3}$ -- 5d $^{5}{\rm D}_{2}^{\rm o}$ & 6756.851 & 7.870 & $-1.760$ & 7.59 & $-3.86$ & ($-7.13$) \\
        &       & 8 &  4p $^{5}{\rm P}_{3}$ -- 5d $^{5}{\rm D}_{3}^{\rm o}$ & 6757.007 & 7.870 & $-0.900$ & 7.59 & $-3.86$ & ($-7.13$) \\
        &       & 8 &  4p $^{5}{\rm P}_{3}$ -- 5d $^{5}{\rm D}_{4}^{\rm o}$ & 6757.171 & 7.870 & $-0.310$ & 7.59 & $-3.86$ & ($-7.13$) \\
\hline
Zn~{\sc i} &  6362 & 6 &  4p $^{1}{\rm P}_{1}^{\rm o}$ -- 4d $^{1}{\rm D}_{2}$ & 6362.338 & 5.796 & $+0.150$ & (7.74) & ($-5.71$) & ($-7.45$) \\
\hline
\end{tabular}
\end{center}
All data were taken from Kurucz and Bell's (1995) compilation
as far as available. RMT is the multiplet number given in the 
Revised Multiplet Table (Moore 1959).
Gammar is the radiation damping constant, $\log\gamma_{\rm rad}$.
Gammas is the Stark damping width per electron density
at $10^{4}$ K, $\log(\gamma_{\rm e}/N_{\rm e})$.
Gammaw is the van der Waals damping width per hydrogen density
at $10^{4}$ K, $\log(\gamma_{\rm w}/N_{\rm H})$. 
Note that the values in parentheses are the default damping parameters 
computed within the Kurucz's WIDTH program (cf. Leusin, Topil'skaya 1987),
because of being unavailable in Kurucz and Bell (1995).
The meanings of other columns are self-explanatory.

\end{table}

\setcounter{table}{2}
\begin{table}[h]
\caption{Abundance variations in response to changing atmospheric parameters.}
\begin{center}
\footnotesize
\begin{tabular}{ccccccc}\hline\hline
Line                  & $\Delta_{T+}$    & $\Delta_{T-}$    & $\Delta_{g+}$    & $\Delta_{g-}$    & $\Delta_{v+}$     & $\Delta_{v-}$    \\
\hline
 S~{\sc i} 8695       & $-0.016$ (0.005)  & $+0.016$ (0.006) & $-0.123$ (0.013) & $+0.135$ (0.015) & $+0.090$ (0.005) & $-0.091$ (0.004) \\
 S~{\sc i} 6757       & $-0.006$ (0.001)  & $+0.007$ (0.001) & $-0.092$ (0.010) & $+0.103$ (0.013) & $+0.079$ (0.003) & $-0.079$ (0.003) \\
\hline
 Zn~{\sc i} 6362       & $-0.022$ (0.005)  & $+0.025$ (0.005) & $-0.047$ (0.019) & $+0.060$ (0.020) & $+0.067$ (0.009) & $-0.065$ (0.010) \\
\hline
\end{tabular}
\end{center}
\footnotesize
Note. Changes of the abundances (expressed in dex) derived from each line 
in response to varying $T_{\rm eff}$ by $\pm 100$~K, $\log g$ by $\pm 0.2$~dex,
and $v_{\rm t}$ by $\pm 0.2$~km~s$^{-1}$. Shown are the mean values averaged over
each of the 239 stars, while those in parentheses are the standard deviations.
\end{table}

\newpage

\setcounter{figure}{0}
\begin{figure}
  \begin{center}
    \FigureFile(160mm,200mm){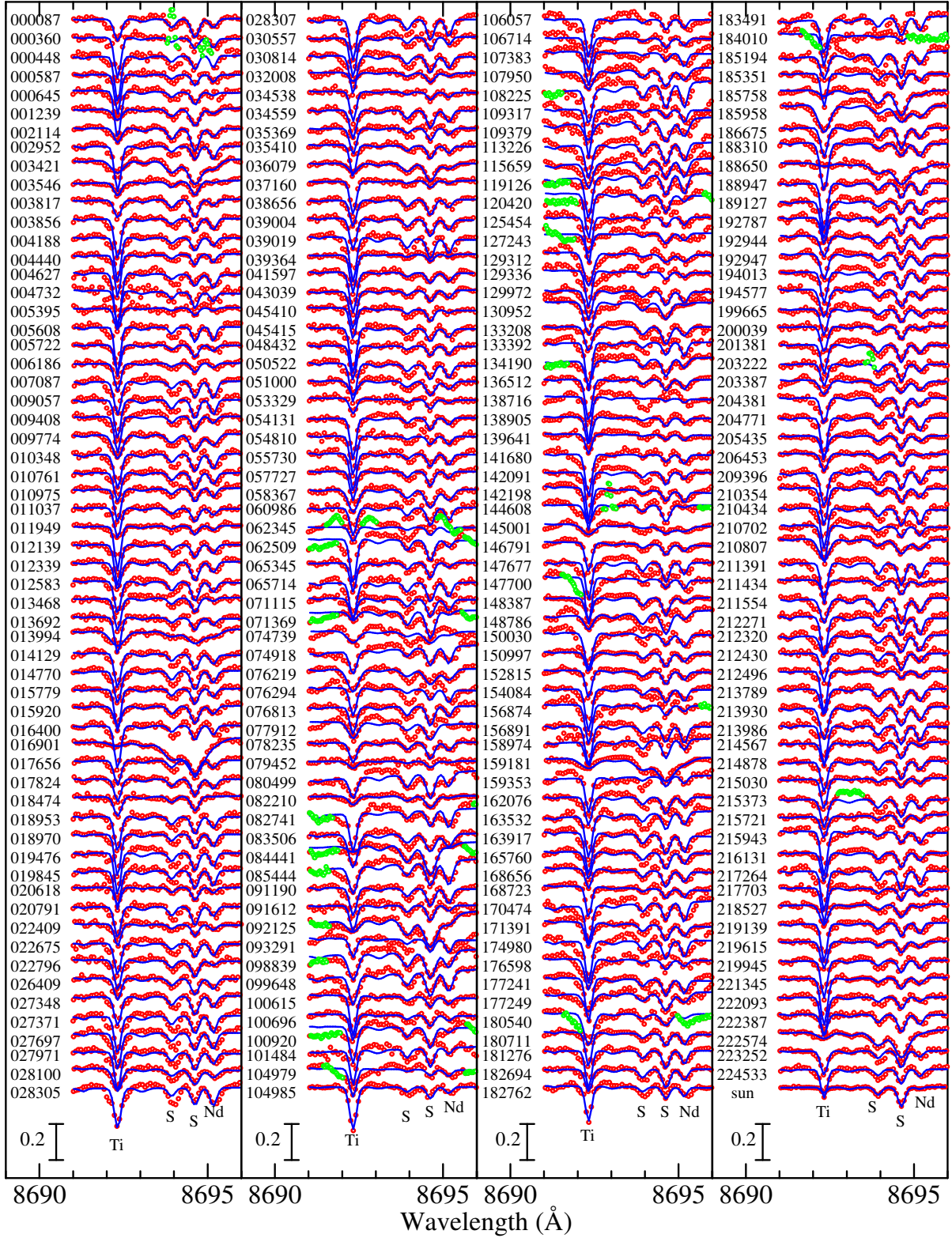}
  \end{center}
\caption{
Synthetic spectrum fitting for 239 GK giants in the 8691--8696~$\rm\AA$ region 
comprising the S~{\sc i} 8693--5 lines. The best-fit theoretical spectra
are shown by blue solid lines, and the observed data are plotted
by red open circles (while those masked/disregarded in the fitting 
are highlighted in green). A vertical offset of 0.1 (in terms of the 
normalized flux with respect to the continuum) is applied to each spectrum
relative to the adjacent ones. Each of the spectra are arranged 
in the increasing order of HD number (indicated on the left to each 
spectrum). The wavelength scale of each spectrum is adjusted to 
the laboratory system.
}
\end{figure}

\setcounter{figure}{1}
\begin{figure}
  \begin{center}
    \FigureFile(160mm,200mm){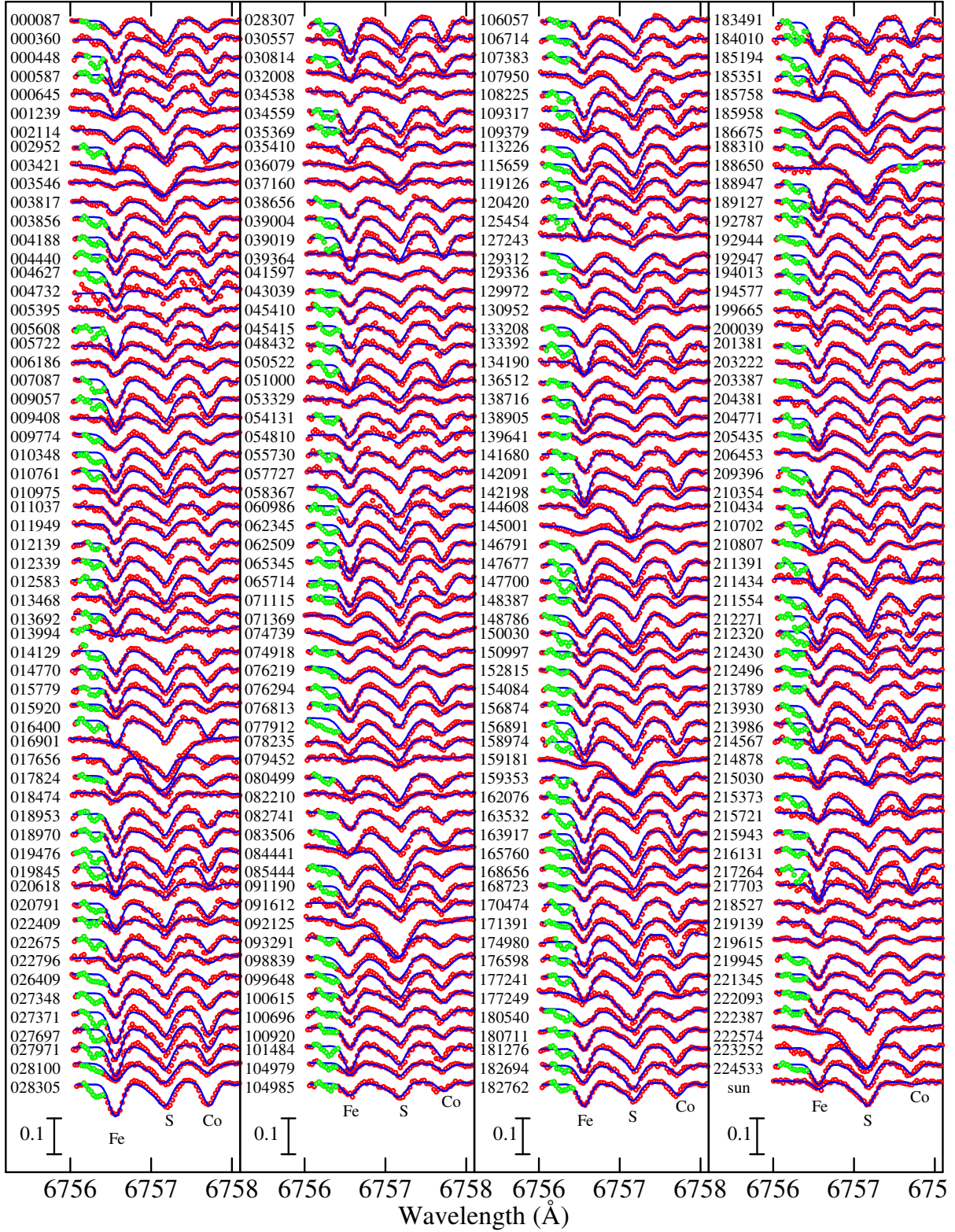}
  \end{center}
\caption{
Synthetic spectrum fitting for 239 GK giants in the 6756--6758.1~$\rm\AA$ region 
comprising the S~{\sc i}~6757 line feature. A vertical offset of 0.05 
is applied to each spectrum. Otherwise, the same as in figure~1.
}
\end{figure}

\setcounter{figure}{2}
\begin{figure}
  \begin{center}
    \FigureFile(160mm,200mm){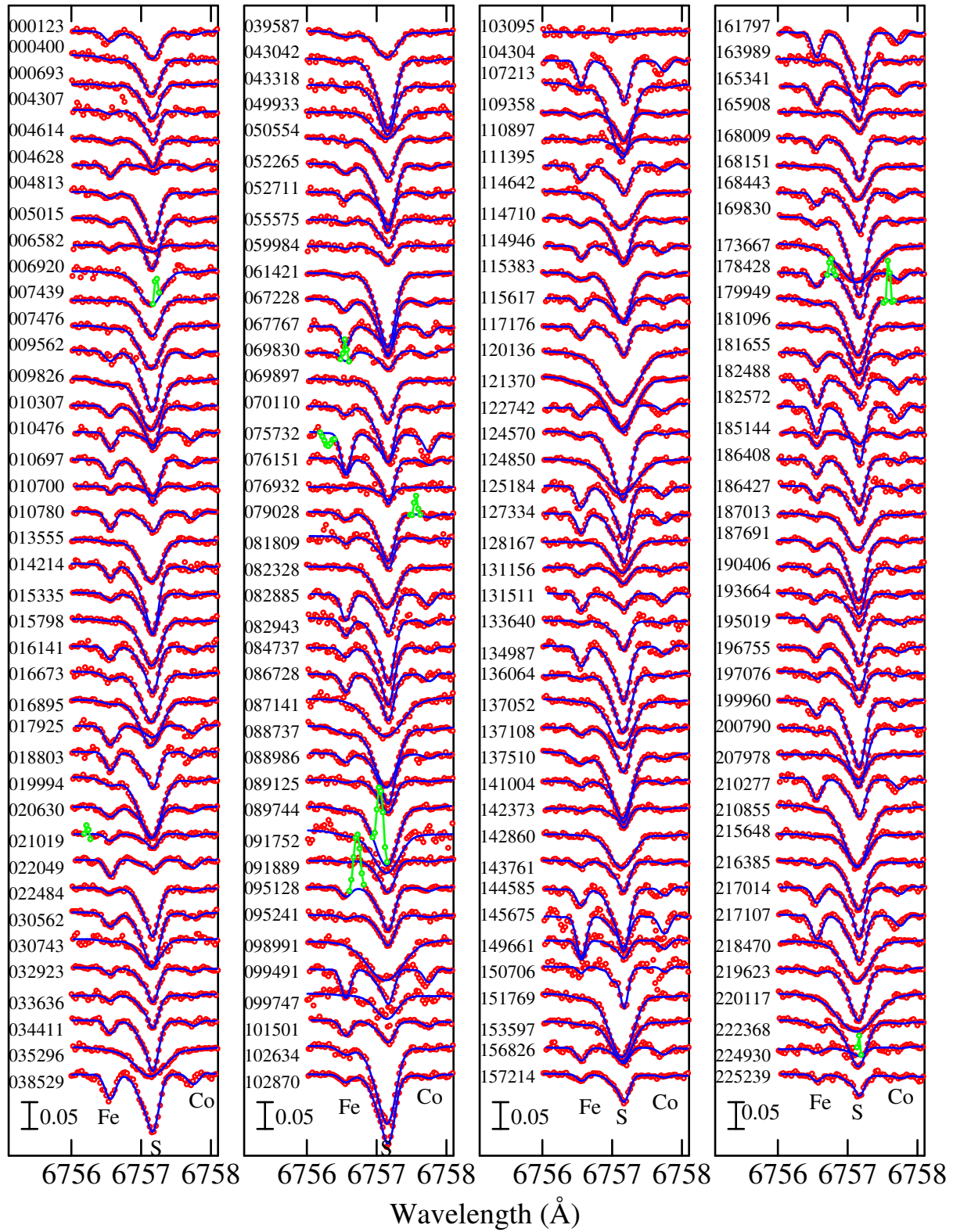}
  \end{center}
\caption{
Synthetic spectrum fitting for 160 FGK dwarfs (and subgiants) in the 
6756--6758.1~$\rm\AA$ region comprising the S~{\sc i} 6757 line feature. 
Otherwise, the same as in figure~1.
}
\end{figure}

\setcounter{figure}{3}
\begin{figure}
  \begin{center}
    \FigureFile(160mm,200mm){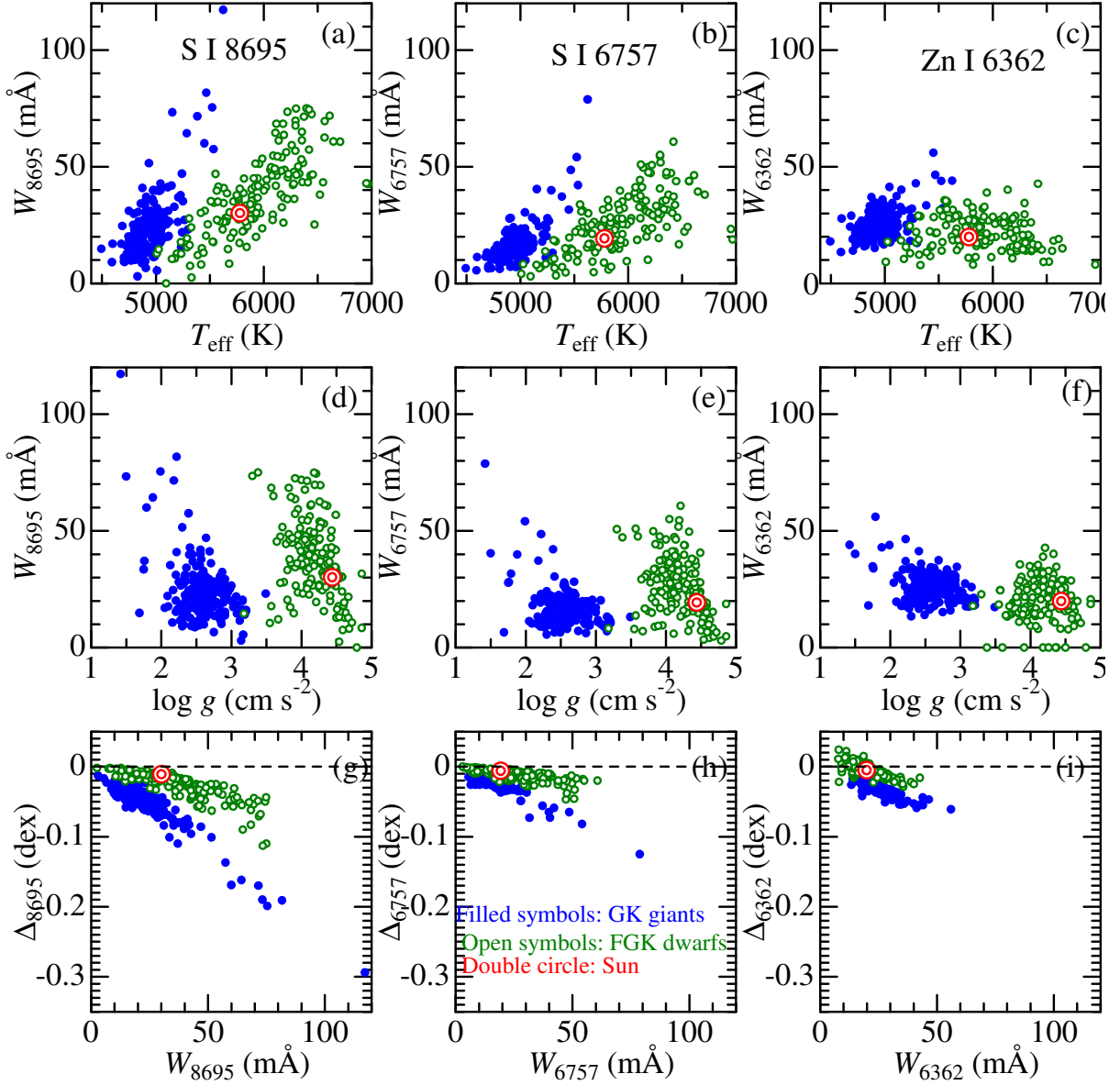}
  \end{center}
\caption{
Equivalent widths ($W$) of S~{\sc i} 8695, S~{\sc i} 6757, and Zn~{\sc i} 6362 lines 
are plotted against the atmospheric parameters in panels (a)--(f), while the dependences
of non-LTE corrections ($\Delta$) upon $W$ are shown in panels (g)--(i):
(a) $W_{8695}$ vs. $T_{\rm eff}$, (b) $W_{6757}$ vs. $T_{\rm eff}$, 
(c) $W_{6362}$ vs. $T_{\rm eff}$, 
(d) $W_{8695}$ vs. $\log g$, (e) $W_{6757}$ vs. $\log g$, (f) $W_{6362}$ vs. $\log g$, 
(g) $\Delta_{8695}^{\rm NLTE}$ vs. $W_{8695}$, (h) $\Delta_{6757}^{\rm NLTE}$ vs. $W_{6757}$, 
and (i) $\Delta_{6362}^{\rm NLTE}$ vs. $W_{6362}$.
The results for 239 GK giants are shown in blue filled symbols, while those for 160 FGK dwarfs
are in open green symbols. In particular, the solar value is indicated by a large double circle
in each panel.
}
\end{figure}

\setcounter{figure}{4}
\begin{figure}
  \begin{center}
    \FigureFile(160mm,200mm){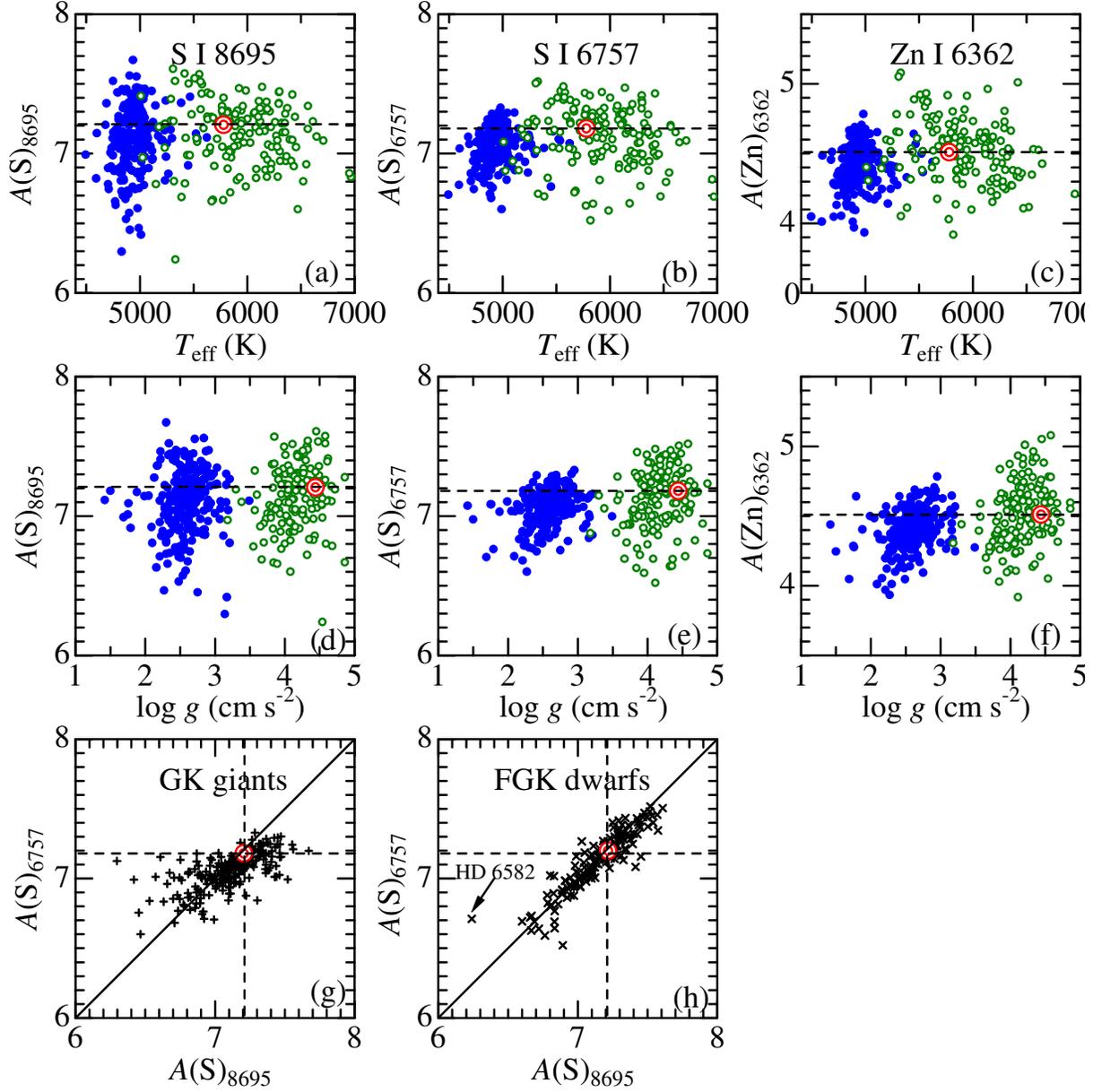}
  \end{center}
\caption{
Non-LTE abundances ($A$) derived from S~{\sc i} 8695, S~{\sc i} 6757, and Zn~{\sc i} 6362 
lines are plotted against the atmospheric parameters in panels (a)--(f), where the meanings 
of the symbols are the same as in figure~4:
(a) $A$(S)$_{8695}$ vs. $T_{\rm eff}$, (b) $A$(S)$_{6757}$ vs. $T_{\rm eff}$, 
(c) $A$(Zn)$_{6362}$ vs. $T_{\rm eff}$, (d) $A$(S)$_{8695}$ vs. $\log g$, 
(e) $A$(S)$_{6757}$ vs. $\log g$, and (f) $A$(Zn)$_{6362}$ vs. $\log g$. 
Panels (g) and (h) show the $A$(S)$_{6757}$ vs. $A$(S)$_{8695}$ correlations 
for GK giants and FGK dwarfs, respectively.
}
\end{figure}

\setcounter{figure}{5}
\begin{figure}
  \begin{center}
    \FigureFile(160mm,200mm){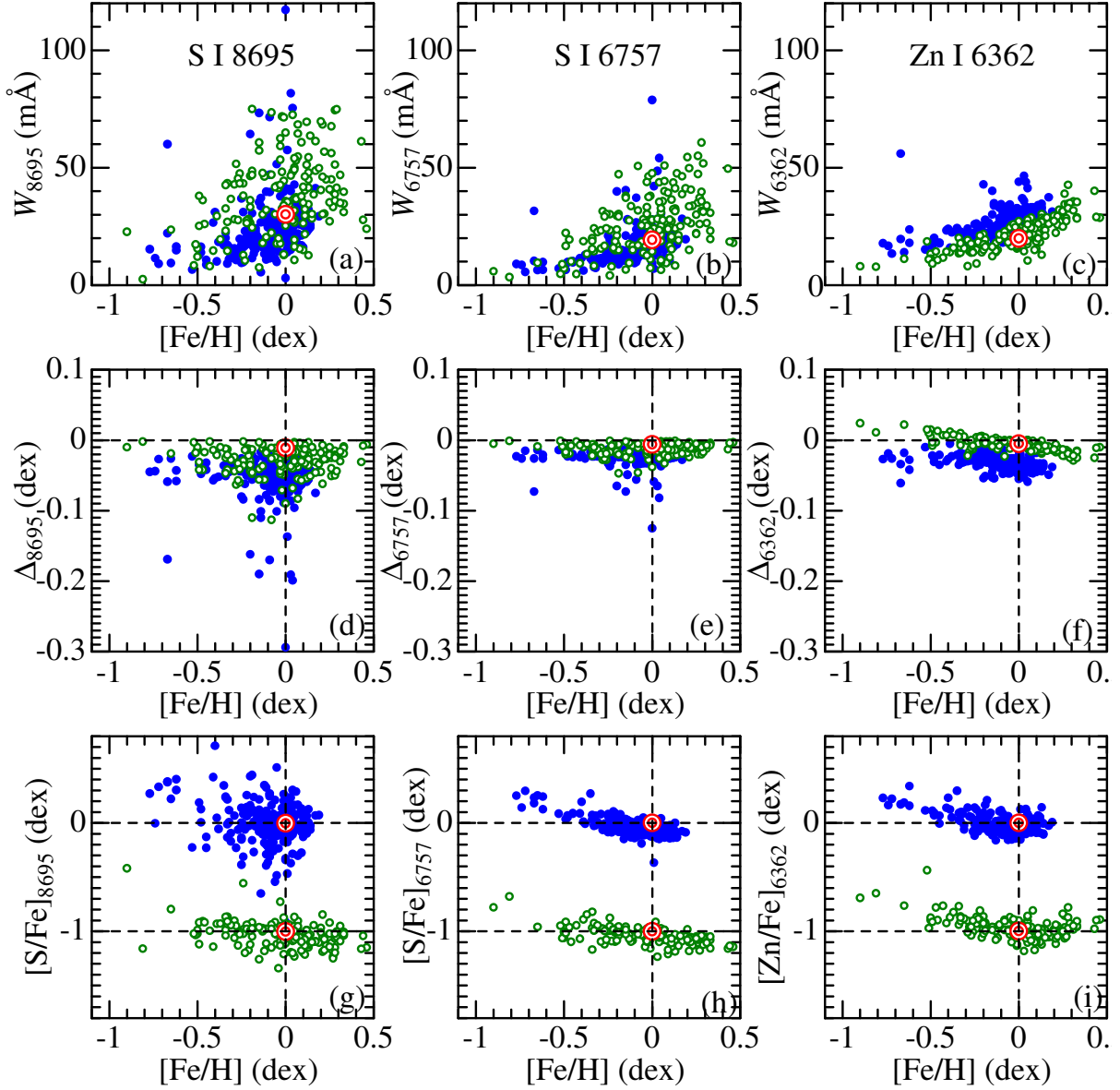}
  \end{center}
\caption{
[Fe/H]-dependences of equivalent widths ($W$), non-LTE corrections ($\Delta$), 
and [X/Fe] ratios (X is S or Zn), derived from S~{\sc i} 8695, S~{\sc i} 6757, 
and Zn~{\sc i} 6362 lines, where the meanings of the symbols are the same as in figure~4:
(a) $W_{8695}$, (b) $W_{6757}$, (c) $W_{6362}$,  
(d) $\Delta_{8695}$, (e) $\Delta_{6757}$, (f) $\Delta_{6362}$,  
(g) [S/Fe]$_{8695}$, (h) [S/Fe]$_{6757}$, and (i) [Zn/Fe]$_{6362}$.
In panels (g)--(h), the results for FGK dwarfs are shown with an offset of
$-1$~dex relative to those of GK giants.
}
\end{figure}

\setcounter{figure}{6}
\begin{figure}
  \begin{center}
    \FigureFile(110mm,160mm){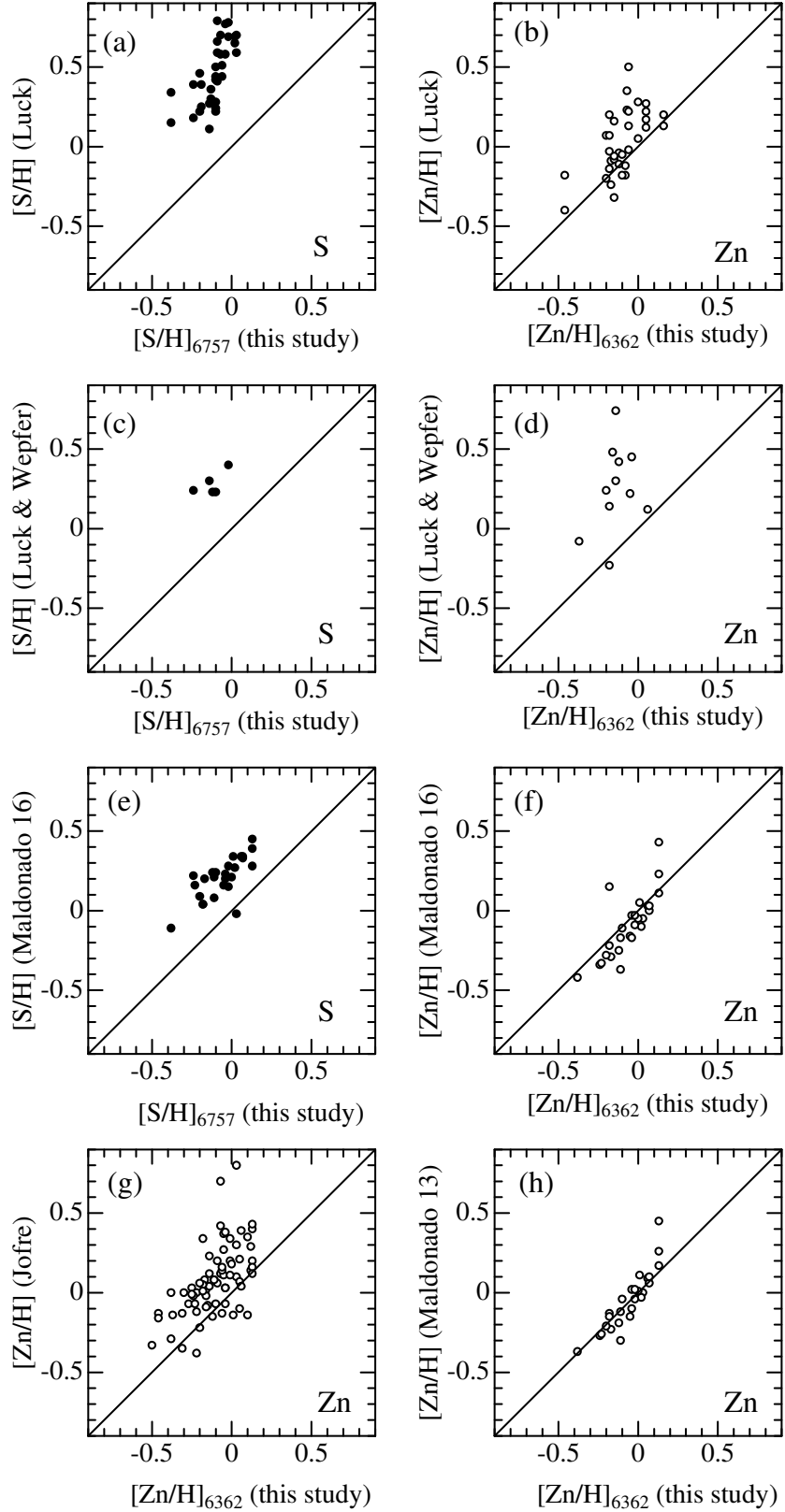}
  \end{center}
\caption{
Comparison of our finally adopted [S/H]$_{6757}$ or [Zn/H]$_{6362}$ results 
of GK giants with those of the common stars available in various literature:
(a) [S/H] from Luck (2015), (b) [Zn/H] from Luck (2015),
(c) [S/H] from Luck \& Wepfer (1995), (d) [Zn/H] from Luck \& Wepfer (1995),
(e) [S/H] from Maldonado \& Villaver (2016), (f) [Zn/H] from Maldonado \& Villaver (2016)
(f) [Zn/H] from Jofr\'{e} et al. (2015), and (g) [Zn/H] from Maldonado et al. (2013).
Filled circles $\cdots$ comparison of [S/H], open circles $\cdots$ comparison of [Zn/H]. 
}
\end{figure}

\setcounter{figure}{7}
\begin{figure}
  \begin{center}
    \FigureFile(120mm,160mm){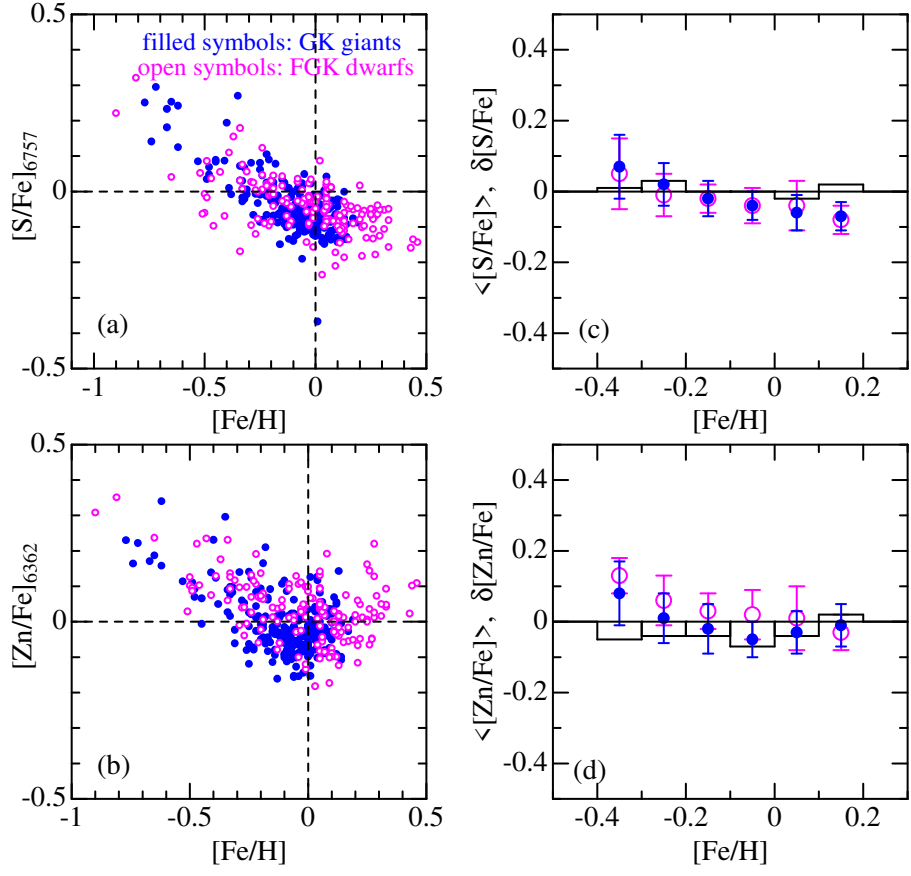}
  \end{center}
\caption{
Left-hand panels (a, b) show the comparison of [S/Fe] (from S~{\sc i} 6757) vs.[Fe/H] 
and [Zn/Fe] (from Zn~{\sc i} 6362) vs. [Fe/H] relations derived in this study for 
239 G--K giants (filled symbols) with those of 160 FGK dwarfs (open symbols).
Symbols in the right-side panels (c, d) give the mean $\langle$[S/Fe]$\rangle$ 
and $\langle$[Zn/Fe]$\rangle$ at each metallicity group (0.1~dex bin within 
$-0.4 \le$~[Fe/H]~$\le +0.2$) where error bars denote the standard deviations, 
while bar graphs represent the mean abundance differences between giants and dwarfs 
defined as $\langle$[X/Fe]$\rangle_{\rm giants}$ $-$ $\langle$[X/Fe]$\rangle_{\rm dwarfs}$. 
}
\end{figure}

\end{document}